\documentstyle[epsf]{aa}
\def\figuredisk{.} 
\def\moverl{(M/L_B)_*}
\begin{document}

\thesaurus{03(11.05.2; 11.19.2; 02.08.1; 03.13.4)}
\title{Star formation and the interstellar medium in low surface
brightness galaxies}
\subtitle{III. Why they are blue, thin and poor in molecular gas}

\author{Jeroen P.E. Gerritsen and W.J.G. de Blok\thanks{\emph{Present address:} School of Physics, University of Melbourne, Parkville, VIC 3052, Australia}}

\institute{Kapteyn Astronomical Institute, 
  Postbus 800, 9700 AV Groningen, The Netherlands}


\date{Received; accepted}

\maketitle
\markboth{Gerritsen \& de Blok: Star formation \& ISM in LSB galaxies}{}

\begin{abstract}
  We present $N$-body simulations of Low Surface Brightness (LSB) galaxies
  and their Interstellar Medium to investigate the cause for their low
  star formation rates (SFR).

  Due to their massive halos, stellar disks of LSB galaxies are very
  stable and thin. Lack of dust makes the projected edge-on surface
  brightness of LSB galaxies comparable to the projected edge-on
  surface brightness of dust-rich High Surface Brightness (HSB)
  galaxies of similar size.

  We show that the low surface densities found in LSB galaxies are by
  themselves not enough to explain the slow evolution of LSB galaxies.
  A low metal content of the gas is essential.  As a consequence the
  gas cools inefficiently, resulting in an almost negligible cold gas
  fraction. We show that LSB galaxies must have molecular gas
  fractions of less than 5 percent.

  Our best model has a SFR which is on average low but fluctuates
  strongly. This causes the large spread in colors of LSB galaxies.
  From a distribution of birthrate parameters we conclude that the
  presently-known and modeled gas-rich blue LSB galaxies constitute
  the majority of the total population of gas-rich LSB disk galaxies.
  We deduce the existence of an additional red, quiescent and gas-rich
  population which constitutes less than 20 percent of the total
  population.  This does not rule out the existence of a large number
  of gas-poor LSB galaxies. These must however have had an
  evolutionary history dramatically different from that of the
  gas-rich galaxies.

\end{abstract}

\keywords{galaxies: evolution --- galaxies: spiral
  --- hydrodynamics --- methods: numerical}

\section{Introduction}

Deep surveys of the night-sky have uncovered a large population of
disk galaxies with properties quite different from those of the
extensively studied ``normal'' high surface brightness (HSB) galaxies.
These so-called Low Surface Brightness (LSB) galaxies have, as their
name already implies, surface brightnesses much lower than what was
until quite recently assumed to be representative for disk galaxies
(Freeman 1970).

The LSB galaxies we will be discussing are generally dominated by an
exponential disk, with scale lengths of a few kpc. Morphologically
they form an extension of the Hubble sequence towards very
late-type galaxies. Observations suggest that LSB galaxies are
unevolved galaxies, as shown by e.g. the sub-solar metallicities
(McGaugh 1994) and the colors (McGaugh \& Bothun 1994; de Blok et
al.\ 1995; van den Hoek et al.\ 1997).

The evolutionary rate of a galaxy may in fact be reflected in its
surface brightness. For example, the gas fraction ($M_{\rm gas} /
M_{\rm gas+stars}$) increases systematically with surface brightness,
from a few percent for early type spirals to much higher values
approaching unity for late type LSB galaxies (\cite{dbm96}). In many
LSB galaxies the gas mass exceeds the stellar mass (even for such
extreme assumptions as maximum disk).

It is still unclear what the physical driver is for the difference
between HSB and LSB galaxies. Investigations of their dynamics, using
H\,{\sc i} observations (de Blok et al.\ 1996), suggest that LSB
galaxies are low-density galaxies (de Blok \& McGaugh 1996).  This is
one of the favored explanations for the low evolution rate of LSB
galaxies (see e.g.\ van der Hulst et al.\ 1987), as this implies a
large dynamical time-scale. Environment may also play a role. Tidal
interactions with other galaxies increase star formation rates. This
may not happen in LSB galaxies which are found to be isolated (Mo et
al. 1994). In LSB galaxies a few star forming regions are usually found.
These are distributed randomly over the galaxies and do not trace the
spiral arms.

The main purpose of this paper is to investigate whether the low
density alone can explain the properties of LSB galaxies. We use 3D
numerical simulations to address this problem. The dynamical
simulations include both stars and gas and incorporate a parameterized
description of star formation, including feedback on the gas. 

The prescription of star formation in numerical simulations is not
straightforward, given the limited resolution of the models and the
fragmentary knowledge about the physics governing star formation on
kpc scales.  In the literature various different star formation
algorithms can be found. For instance, Friedli \& Benz (1995) use a
criterion based on the Toomre stability parameter $Q$ (Toomre 1964) to
study star formation in barred systems; Mihos \& Hernquist (1994a,b, 
1996) adopt a Schmidt law based on the gas density [star
formation rate (SFR) $\propto\rho^{1.5}$] to study mergers of
galaxies; others require gas to be in a convergent and Jeans
unstable flow to form stars (\cite{kat92}; \cite{kat96}; 
\cite{nav93}; \cite{ste96}).

We employ the method of Gerritsen \& Icke (1997, 1998), which uses a
Jeans criterion to define star forming regions, coupled with an
estimate of the cloud collapse time. In these simulations gas is
treated fully radiative with allowed temperatures between 10 K and
$10^7$ K; cooling is described by standard cooling functions, heating
is assumed to be provided by far-ultraviolet (FUV) radiation and
mechanical heating from stars. In this way the simulated interstellar
medium (ISM) mimics a multi-phase ISM (\cite{fie69}; \cite{mck77}).
This is an improvement over other simulations which try to create a
multi-phase ISM but do not allow radiative cooling below $10^4$ K
(e.g.\ \cite{her89a}; \cite{kat96}).  The multi-phase ISM allows us to
restrict ourselves to considering only relatively cold ($T<10^3$ K)
regions as the sites for star formation.

Here we apply this method to study the ISM and star forming properties
of LSB galaxies, and our results will be valid under the assumption
that the physical processes governing star formation are the same for
HSB and LSB galaxies.  We test whether a low mass-density is
sufficient to explain the properties of LSB galaxies by discussing two
different models of a specific LSB galaxy: the first model has the
cooling properties of a solar abundance ISM, while in the second model
the cooling efficiency is lowered, thus mimicking a metal-poor ISM.


The structure of the paper is as follows.  In Sect.\ \ref{numtech} we
describe the numerical techniques.  The construction of a model
LSB galaxy is presented in Sect.\ \ref{galmod}.  This galaxy model is
evolved for a few Gyr, using two different prescriptions for the
cooling properties of the gas (Sect.\ \ref{evolution}).  The implications of
the simulations are discussed in Sect.\ \ref{discus}.  We conclude this paper
with a summary in Sect.\ \ref{conclusion}.

\section{Numerical technique}\label{numtech} We model the evolution of
galaxies using a hybrid {\it N-}body/hydrodynamics code (TREESPH;
Hern\-quist \& Katz 1989).  An extensive description is given in
Gerritsen \& Icke (1997, 1998). A tree algorithm (\cite{bar86};
\cite{her87}) determines the gravitational forces on the
collision-less and gaseous components of the galaxies. The
hydrodynamic properties of the gas are modeled using smoothed particle
hydrodynamics (SPH) (see \cite{luc77}; \cite{gin77}). The gas evolves
according to hydrodynamic conservation laws, including an artificial
viscosity for an accurate treatment of shocks.  Each particle is
assigned an individual smoothing length, $h$, which determines the
local resolution and an individual time step.  Estimates of the gas
properties are found by smoothing over 32 neighbors within $2h$.  We
adopt an equation of state for an ideal gas.




We allow radiative cooling of the gas according to the cooling
function for a standard hydrogen gas mix with a helium mass fraction
of 0.25 (\cite{dal72}). Radiative heating is modeled as photo-electric
heating of small grains and PAHs by the FUV field (\cite{wol95}),
produced by the stellar distribution.

All our simulations are advanced in time steps of $3\times10^6$ yr for
star particles, while the time steps for SPH particles can be 8 times
shorter ($3.8 \times 10^5$ yr). The gravitational softening length is
150 pc. A tolerance parameter $\theta=0.6$ is used for the force
calculation, which includes quadrupole moments.

\subsection{Star formation and feedback}\label{starform} 
Gerritsen \& Icke (1997, 1998) extensively describe the recipe for
transforming gas into stars and the method for supplying feedback onto
the gas. The recipe works well for normal HSB galaxies, with the
energy budget of the ISM as prime driver for the star formation.  The
simulations allow for a multi-phase ISM with temperature between
$10<T<10^7$ K. This allows us to consider cold $T<10^3$ K regions as
places for star formation (Giant Molecular Clouds in real life).
Below we summarize the important ingredients of the method.

From our SPH particle distribution we select conglomerates where the
Jeans mass is below the mass of a typical Giant Molecular Cloud. In 
the simulations performed here we use $M_c=10^6\ M_{\sun}$
for this aggregate mass (the method is not very sensitive to the exact
value for this mass). The Jeans mass depends on the 
local gas properties and is calculated from the SPH estimates of the
density $\rho$ and sound speed $s$,
\begin{equation} 
  \label{mjeans}
  M_J = {1\over6}\pi\rho \left({\pi s^2\over
      G\rho}\right)^{3\over2},
\end{equation}
with $G$ the constant of gravity.  During the simulations the maximum
number density achieved is of the order 1 cm$^{-3}$. It follows that
only regions below $10^3$ K are unstable and may form stars. 

We follow unstable regions during their dynamical and thermal
evolution and if an SPH particle resides in such a region longer than
the collapse time, 
\begin{equation}
  \label{tdyn}
  t_u > t_c={1\over\sqrt{4\pi G\rho}},
\end{equation}
half of its mass is converted into a star particle.
Experiments with a different number of particles and different star
formatin efficiencies show no dependence on these parameters, as
already shown in Gerritse \& Icke (1997).

Important for our calculations is that we consider star particles as
stellar clusters with an age. Thus for each individual star particle
we can attribute quantities like the SN-rate, the mass loss, and the
FUV-flux, according to its age. We use the spectral synthesis models
of Bruzual \& Charlot (1993) to determine these quantities, where we
adopt a Salpeter Initial Mass Function (IMF) with slope 1.35 and with
lower and upper mass limits of $0.1\ M_{\sun}$ and $125\ M_{\sun}$
respectively.

The radiative heating for a gas particle is calculated by adding the
FUV-flux contributions from all stars, which is done together with the
force calculation.  The mechanical luminosity from a star particle is
determined by both the SN-rate and the mass loss rate.  We assume that
each SN injects $10^{51}$ ergs of energy and that the energy injected
by stellar winds is $E_{w}={1\over2}\Delta m_{\ast} v_{\infty}^2$,
with $v_{\infty}$ the wind terminal velocity and $\Delta m_{\ast}$ the
stellar mass loss.  For massive stars $v_{\infty}$ depends critically
on the stellar mass, luminosity, effective temperature, and
metallicity (e.g.\ Leitherer et al.\ 1992; \cite{lam93}). For simplicity we
adopt $v_{\infty}=2000$ km/s. After $3.3\times10^7$ yr the last $8\ 
M_{\sun}$ stars explode and no more mechanical energy is supplied to
the gas. Thus we return mechanical energy from massive stars into the
ISM, and ignore the mechanical energy from low mass stars.

In the simulations the parent SPH particle is the carrier of the
mechanical energy from the new star particle. This SPH particle (``SN
particle'') mimics a hot bubble interior. Radiative cooling is
temporarily switched off (the resolution does not allow the creation
of a low-density, hot bubble), the temperature of the SN particle is
set to the mechanical energy density (of a few $10^6$ K), and the
particle evolves adiabatically. The first $10^7$ yr, the position and
velocity of the SN particle are equal to the position and velocity of
the associated star particle. Afterwards, the SN particle evolves
freely. After $3\times10^7$ yr radiative cooling is switched on again,
and the particle behaves like an ordinary SPH particle.

\section{Modeling an LSB galaxy}\label{galmod}

There are many ways to construct model galaxies. For our purpose we
prefer to build a galaxy model after an existing galaxy with
well-determined properties. Here we choose LSB galaxy F563-1; this
galaxy is a late-type LSB galaxy, representative of the field LSB
galaxies found in the survey by Schombert et al.\ (1992).  The
optical properties of this galaxy are described in de Blok et al.\ 
(1995); measurements of metallicities in H\,{\sc ii} regions are
described in de Blok \& van der Hulst (1998a); a neutral hydrogen map
and rotation curve are given in de Blok et al.\ (1996). Parameters
such as stellar velocity dispersion, which cannot be measured
directly, are set in comparison with values measured locally in the
Galaxy. The current star formation rate as deduced from H$\alpha$
imaging is taken from van den Hoek et al.\ (1997).  For convenience
these data are summarized in Table \ref{lsbpar}.

\begin{table}
  \caption[]{Parameters for F563-1 (H$_0$ = 75 km s$^{-1}$
    Mpc$^{-1}$).} 
  \label{lsbpar}
  \begin{flushleft}
    \begin{tabular}{lll}
      \hline\noalign{\smallskip}
      $L_B$ & $1.35\times 10^9\ L_{\sun}$ \\
      $h_*$ & 2.8 kpc & center ($R<5$ kpc)\\
      $h_*$ & 5.0 kpc & outside ($R>5$ kpc)\\
      SFR & 0.05 $M_{\sun}$/yr \\
      \noalign{\smallskip}
      $M_{\rm H\,\sc I}$ & $2.75\times 10^9\ M_{\sun}$ \\
      $v_{\rm max}$ & 113 km/s \\
      \noalign{\smallskip}
      $\rho^{\rm halo}_{0}$ & 0.0751 $M_{\sun}/{\rm pc}^3$ \\
      $R^{\rm halo}_c$ & 1.776 kpc \\
      \noalign{\smallskip}
      \hline\noalign{\smallskip}
    \end{tabular}
  \end{flushleft}
\end{table}

The most difficult problem we face in constructing a model is
converting the measured luminosity to a stellar disk mass. This is one
of the most persistent problems in analyzing the dynamics of galaxies,
and, unfortunately, the present observations do not provide a unique
answer for this stellar disk mass-to-light ratio $\moverl$.  Rather
than using the so-called ``maximum disk'' value $\moverl= 9$, which is
an upper limit to the possible values of $\moverl$, we adopt a value
based on colors and velocity dispersions of $\moverl=1.75$.  An
extensive motivation for this choice is given by de Blok \& McGaugh
(1997).  The implications of choosing a different value of $\moverl$
for the evolution of the stellar disk will be discussed in Sect.\
\ref{sfhist}.

\subsection{The model}\label{modelletje}
\begin{figure}
  \epsfysize=\hsize \hfil\epsfbox{\figuredisk/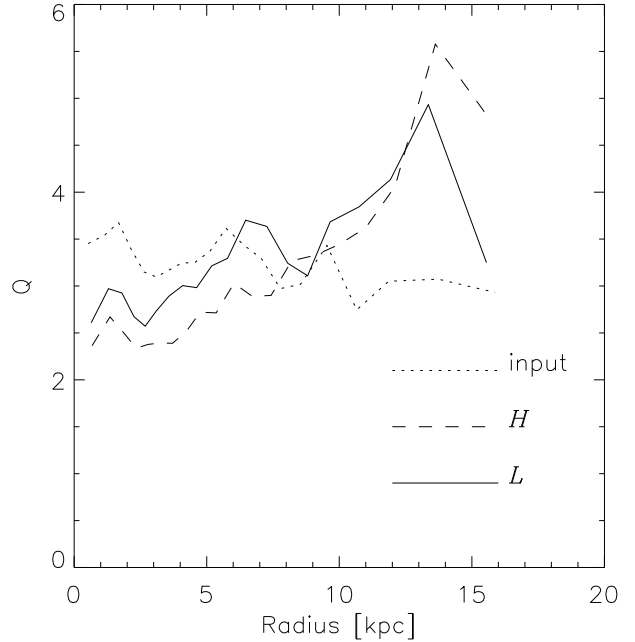}\hfil
  \caption{The stability parameter Q versus radius. The dotted line shows
    the Q-value upon initialization of the galaxy model. The solid
    line denotes simulation $L$, the dashed line simulation $H$. Both
    simulation yield approximately the same stability for the stellar
    disk \label{qtoomre}}
\end{figure}

The stellar disk particles are distributed radially according to the
(measured) surface density profile. We adopt a vertical distribution
of the form sech$^2(z/z_{\ast})$ (e.g.\ van der Kruit \& Searle 1982), with a constant
vertical scale height $z_{\ast}=0.5$ kpc The disk is truncated at 25
kpc. The luminosity yields a total stellar mass of
$M_{\ast}=2.36\times10^9\ M_{\sun}$.  The velocity dispersion of the
stars is fixed via the relation
\begin{equation}
  \label{dispsurf}
  \sigma_z=\sqrt{\pi G z_{\ast}\Sigma_{\ast}},
\end{equation}
where $\Sigma_{\ast}$ is the stellar surface density.
This implies a stability parameter $3<Q<3.5$
throughout for the disk (dotted line in Fig.\ \ref{qtoomre}).

Particle velocities are assigned according to the (gas)
rotation curve corrected for asymmetric drift. Dispersions in $R, z,
\theta$ directions are drawn from Gaussian distributions with
dispersions of $\sigma_R, \sigma_z, \sigma_{\theta}$ respectively,
using the relations valid for the solar neighborhood
\begin{equation}
  \label{disprel}
  \sigma_z=0.6\sigma_R,\ \ 
  \sigma_{\theta}={\kappa\over 2\omega}\sigma_R,
\end{equation}
with $\omega$ and $\kappa$ the orbital and epicyclic frequencies
respectively.

The gas surface density distribution is modeled after the H\,{\sc i}
distribution. The total gas mass is $M_g=3.85\times10^9\ M_{\sun}$
(this is the total H\,{\sc i} mass multiplied by 1.4 to include He);
the surface density decreases almost linearly with radius out to 33
kpc.  The vertical distribution is assumed to decline exponentially.
The scale height $z_g$ of the gas layer can be calculated using
\begin{equation}
  \label{veldis}
  \sigma_g^2 = 2\pi G(\Sigma_g+\Sigma_{\ast}){z_g^2\over
  {1\over2}z_{\ast}+z_g},
\end{equation}
(\cite{dop94}) where $\sigma_g=3$ km/s is the (adopted) gas velocity
dispersion.  The gas particles are assigned velocities according to
the rotation curve, with isotropic dispersion $\sigma_g$.

The halo is included in the calculations as a rigid potential. This is
justified since the galaxy model evolves in isolation. We will thus
also ignore any contraction of the halo under the influence of the
disk potential. We do not expect this effect to be important anyway as
the mass of the disk (assuming $\moverl=1.75$) is only 4 per
cent of the measured halo mass (de Blok \& McGaugh 1997).  As
advantages we do not have to make assumptions about halo particle
orbits, and we do not have to spend time in calculating the force of
the halo particles on the galaxy. We assign an isothermal density
distribution to the halo,
\begin{equation}
  \label{halo}
  \rho_h={\rho_0\over 1+(r/R_c)^2},
\end{equation}
with central volume density $\rho_0=0.0751 M_{\sun}/{\rm pc}^3$ and
core radius $R_c=1.776$ kpc. These are the values derived from a
rotation curve decomposition assuming $\moverl = 1.75$. This
will correctly put the maximum rotation velocity at 113 km/s.

For the simulations we use 40,000 SPH particles and 80,000 star
particles initially. This corresponds to an SPH particle mass of
$9.6\times10^4\ M_{\sun}$ and a star particle mass of
$3.0\times10^4\ M_{\sun}$.

\subsection{Implications of $\moverl$ for star formation history} 
\label{sfhist}

We can show that a maximum disk value for $\moverl$ is not a plausible
option for our modeling exercise, and that in fact the ``most
likely'' value we use in constructing models may still overestimate
the true value of $\moverl$, making the stellar disk a really insignificant
component of the whole galaxy system.

The star formation history of
galaxies is commonly parameterized by an exponential function:
\begin{equation}
  \label{ages}
  {\rm SFR}=-\dot M_g = M_g/ \tau_{\ast}
\end{equation}
where $\tau_{\ast}$ is the star formation time scale (e.g.\ Guiderdoni
\& Rocca-Volmerange 1987 and \cite{cha91}). This ignores any gas
locked up in long-lived stars, but this effect will not be important
in LSB galaxies due to their large gas-fractions. A small value for
$\tau_{\ast}$ means that star formation proceeds rapidly, while a
large $\tau_{\ast}$ indicates that star formation proceeds very
slowly.

For HSB galaxies the star formation time scale is normally a few Gyr
(\cite{ken94}). For F563-1 the measured current SFR as derived from
H$\alpha$ imaging is approximately $0.05\ M_{\sun}$/yr; this yields a
star formation time scale of 77 Gyr. Thus star formation in this disk
proceeds much slower than in HSB galaxies.

Since we know the stellar and gas mass, and hence the total disk mass
$M_t=M_*+M_g$ of the galaxy, we can use the exponential
parameterization and the known SFR to compute the ``age'' $t_d$ of the
disk (i.e.\ the time elapsed since star formation started):

\begin{equation}
  \label{mstar}
  M_* = -\int_0^{t_d} \dot M_g dt = M_t \left( 1-{\rm
  e}^{-t_d/\tau_*} \right).
\end{equation}
For the adopted stellar mass of F563-1 Eq.\ \ref{mstar} yields an age
for the galaxy of 37 Gyr, much larger than the age of the universe.

If the exponentially declining SFR describes the star formation
history adequately and the galaxy is not in a state of unusually low
star formation activity (note that the blue colors for these LSB
galaxies suggest a relatively {\it high\/} current star formation
activity), then the old implied age of the disk could mean that the
mass of the disk should in reality be even less.

\begin{figure}
  \epsfysize=\hsize \hfil\epsfbox{\figuredisk/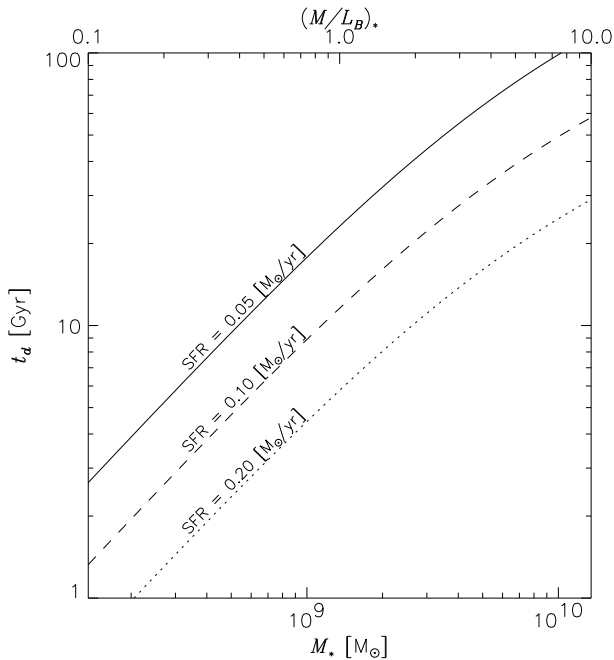}\hfil
  \caption{Relation between galaxy age, $t_d$ (vertical axis), and
  stellar mass, $M_*$ (horizontal axis), according to Eqs.\ \ref{ages}
  \& \ref{mstar}. The different lines are solutions to these equations
  for various values of the current SFR, where the measured value is
  $0.05\ M_{\sun}$/yr. The top axis shows the $\moverl$ ratio
  corresponding to the stellar mass. The input model assumes
  $\moverl=1.75$, the maximum disk model has $\moverl=9$, which yields
  a very high age for the galaxy \label{m2l}}
\end{figure}

We show this in Fig.\ \ref{m2l} where we present a few solutions for
Eqs.\ \ref{ages} \& \ref{mstar} for various values of the SFR. If we
adopt $0.05\ M_{\sun}$/yr as the true SFR then $\moverl\approx0.6$ if
the galaxy is 15 Gyr old. Higher values for $\moverl$ would make the
galaxy older than the universe. A maximum disk solution (\cite{val86}) for the
stellar mass-to-light ratio  yields $\moverl=9$, which
leads to an age of 110 Gyr. A current SFR of order $0.4\ M_{\sun}$/yr
is needed to reconcile the maximum disk with the exponentially
declining SFR model, which implies that the measurements underestimate
the true SFR by an order of magnitude. Thus seems highly unlikely (it
would give LSB galaxies SFRs comparable to those of actively star
forming late type galaxies), and therefore we reject the maximum disk
as an acceptable solution for the stellar disk mass.

The adopted $\moverl=1.75$ which we use in the galaxy model is only
consistent with the age of the universe if the true SFR is of order
$0.10\ M_{\sun}$/yr, otherwise the stellar mass is probably
overestimated. However if the stellar disk is indeed less massive than
our galaxy model, then any conclusions concerning the disk stability
and thickness of the disk will be even stronger.

It might be argued that the exponential star formation history is not
the right choice for this type of galaxy, and that other functional
forms of star formation history, in combination with the measurements,
will yield reasonable ages. This is not true, as any other
(reasonable) star formation history will result in even larger ages
for the disk.  Taking for example another extreme functionality -- a
constant SFR of 0.05 $M_{\sun}$/yr -- yields an age $t_d = 47$
Gyr. The only way to get a reasonable age, given the measured
luminosity and assumed $\moverl$, is to assume that the current
SFR in F563-1 (and LSB galaxies in general) has been underestimated,
which is unlikely. We refer to McGaugh \& de Blok (1997) for an
extensive discussion on the various functionalities of the star
formation history.

We therefore conclude that the disk of F563-1 must have a small
stellar mass-to-light ratio.

\section{Evolution} \label{evolution}

In this section we demonstrate that the low density as found in LSB
galaxies, by itself is not sufficient to reproduce the low observed
SFRs of LSB galaxies. {\it Low metallicity gas is required to explain
  the properties of LSB galaxies.}

To show this we construct two model galaxies using the structural
parameters of F563-1.  Model $H$ represents an LSB galaxy with a solar
metallicity gas.  Although we use the structural parameters relevant
for F563-1, the model is in effect a model HSB galaxy,
which is ``stretched out'' to give the low (surface) densities found
in LSB galaxies. This model therefore tests the low-density
hypothesis.

The other model, $L$, has the same structural parameters as $H$, but
in addition we lowered the cooling efficiency of the gas below $10^4$
K by a factor of seven. Cooling below $10^4$ K is dominated by metals,
so lowering the efficiency is equivalent to lowering the metallicity
by an equal amount.  Model $L$ thus most closely approximates what is
currently known observationally about LSB galaxies. In summary we
adopt the cooling function of a solar abundance gas in model $H$.  This
cooling function is later changed in model $L$ to simulate the effects
of low metallicity.

We assume that the physics regulating the star formation is the same
for LSB and HSB galaxies, and therefore use the star formation recipe
described in Section 2 and applied to HSB galaxies in Gerritsen \&
Icke (1997, 1998).  After initialization we let the two model galaxies
evolve for 2.3 Gyr. The models rapidly settled into equilibrium, 
any longer time interval would have produced identical results.

In this section we first discuss the evolution of the stellar disk.
The stellar disk evolves rather independently from the gas disk and
star formation. Hence we effectively explore the consequences of the
disk/halo decomposition, notably on the stability and thickness of the
stellar disk.  Second, we discuss the evolution of the SFR with time
for both simulations. For a physical interpretation of the difference
in SFRs we present phase diagrams of the ISM in both simulations.

\subsection{Stellar disk}

Figure \ref{qtoomre} shows the stability parameter $Q$ for the stellar
disk at $t=2$ Gyr (that is 2 Gyr after the start of the simulation
which started at $t=0$), where $Q$ is defined as 
\begin{equation}
  \label{qto}
  Q = {\kappa\sigma_R\over 3.36G \Sigma_{\ast}},
\end{equation}
with $\kappa$ the epicycle frequency (\cite{too64}).  $Q<1$ denotes a
(local) instability, while $Q>1$ means stability.  For both
simulations $Q\approx2.5$ in the center and rises to $Q\approx5$ in
the outer parts of the disk. There is no evolution in the stability.
Once the system has settled after the start of the simulation, the
radial behavior of $Q$ as shown is reached. The value of course
depends on the input parameters, notably the stellar mass, but it is
clear that the stellar disks of LSBs are more stable than the stellar
disks of HSBs, where $Q$ is of order 2 (\cite{bot93}, van der Hulst et
al.\ 1993).

In both simulations the scale height of the stellar disk decreases by
approximately a factor of 1.3.  The initial scale height corresponded
to the scale height of an Sc galaxy (van der Kruit \& Searle 1982).  The final axial
ratio for the galaxy model is about 15.  As particle scattering during
the simulation tends to increase the thickness of the disk, we
conclude that the stellar disks of LSB galaxies are thinner than those
of HSB galaxies. We will return to this in more detail in Sect.\ 5.1.

\subsection{Star formation \& ISM}
\begin{figure}
  \epsfysize=\hsize \hfil\epsfbox{\figuredisk/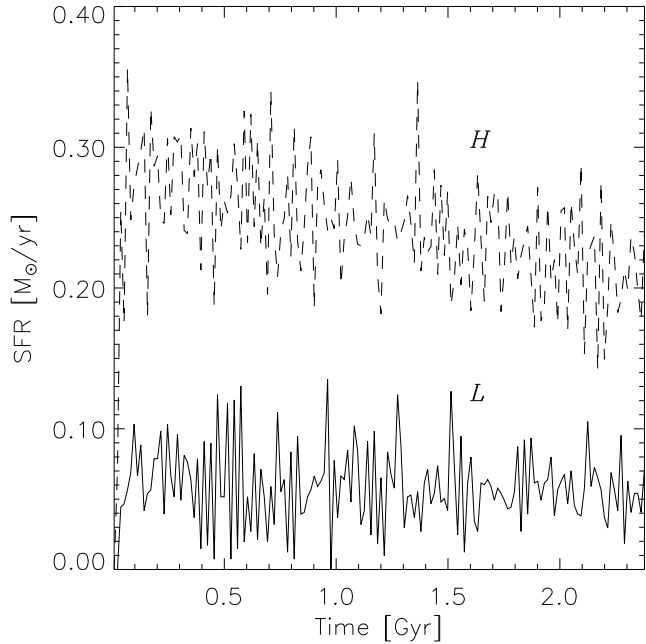}\hfil
  \caption{The evolution of the SFR versus time. The lower line shows
    the SFR from simulation $L$, the upper line represents simulation
    $H$.  Data are averaged over 9 Myr\label{sfrate}}
\end{figure}
Figure \ref{sfrate} shows the SFR versus time.  In model $H$ the SFR
settles after a short adjustment at a value of $0.28\ M_{\sun}$/yr and
declines slowly with time.  The SFR in model $L$ settles at $0.08\ 
M_{\sun}$/yr. 

For both simulations the SFR varies on time scales of a few tens of
Myr and the amplitude of these variations can exceed $0.1\
M_{\sun}$/yr. The rapid variability in the SFR is due to the discrete
nature of star formation in our simulation. New star particles have a
mass of approximately $5 \times 10^4\ M_{\sun}$ and thus represent
(large) stellar clusters. In this way our simulations incorporate the
idea that most stars in real galaxies form in stellar clusters. For a
small number of star forming regions (as in LSB galaxies) this will in
real life also lead to a rapidly fluctuating, but on average low SFR.

In simulation $L$ the strength of the star formation peak is large
compared to the average SFR. These fluctuations in star formation
activity will actually dominate the color of the galaxy, with large
fluctuations giving rise to blue colors. The same star formation
fluctuations will also be present in HSB galaxies, but due to the
higher SFR the large number of fluctuations will give the impression
of a high, relatively constant SFR.

We thus find that it is not so much the absolute value of the average
SFR which determines the colors, but the contrast of any SF
fluctuation with respect to the average SFR.  For the LSB galaxies the
large contrast leads to blue colors.

\begin{figure}
  \epsfxsize=\hsize \hfil\epsfbox{\figuredisk/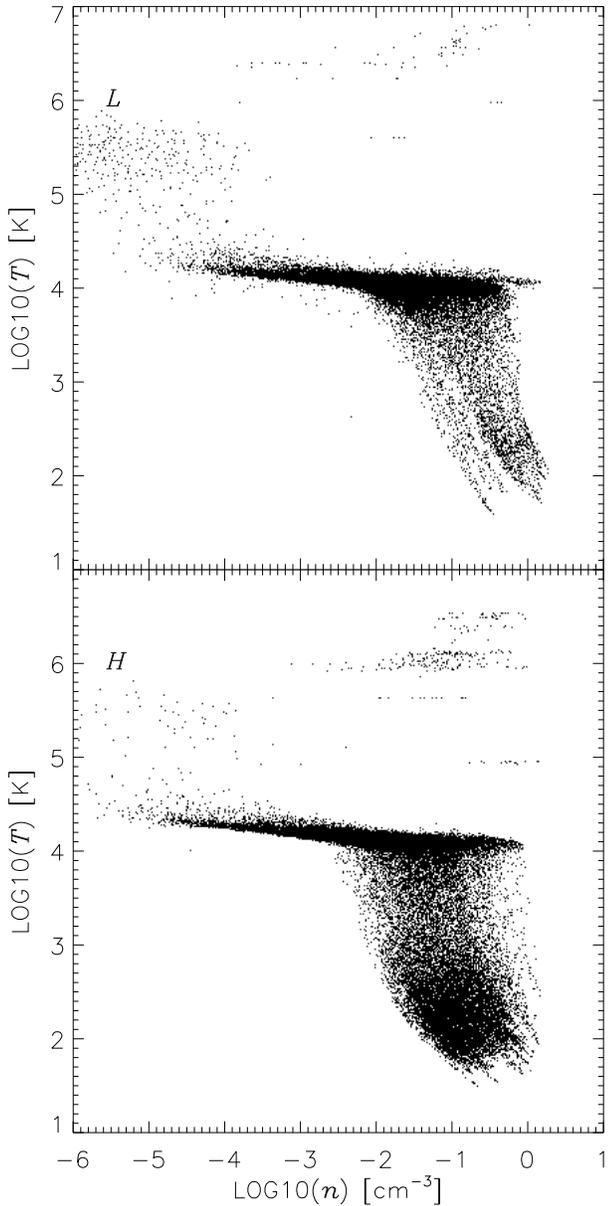}\hfil
  \caption{Phase diagrams (temperature versus number density) for the
    simulations. Top panel shows simulation $L$, bottom panel shows
    simulation $H$; each dot represents an SPH particle. The
    individual particles at the top of each diagram are hot SN
    particles. Simulation $H$ clearly shows a two-phase structure,
    while in simulation $L$ almost all gas is in the warm
    ($T\approx10^4$ K) phase (85\%) \label{ism}}
\end{figure}

In order to understand the behavior of the star formation activity in
the simulations we constructed phase diagrams of the ISM.  Figure
\ref{ism} shows the temperature versus density for all gas particles
in both simulations. The top panel shows simulation $L$ while the
bottom panel shows the phase diagram for simulation $H$. Most
particles have a temperature of $10^4$ K. In simulation $H$ a large
fraction of the particles has a temperature of about 100 K. This
simulation shows a dominant two-phase structure, and resembles the ISM
in simulations of HSB galaxies (\cite{ger97b}).  Quantitatively, the
cold gas fraction ($T<1000$ K) makes up 37\% of the total gas mass in
simulation $H$ and only 4\% in simulation $L$.  This reflects the
cooling properties of the gas: in simulation $L$ seven times less heat
input is required to keep the gas at $10^4$ K as in simulation $H$.

Consequently simulation $L$ contains virtually only a warm, one-phase
ISM. The simulated absence of metals prevents the ISM from cooling
efficiently. As molecular gas is only formed in/from the cold
component of the ISM, the 4\% mentioned above is actually an upper
limit to the amount of molecular gas that could form in such a galaxy.
The amount of cold molecular gas in sites for star formation would
thus be negligible. From these models we expect the disks of LSB
galaxies to contain only negligible amounts of molecular gas. This is
consistent with observations by Schombert et al. (1990) and de Blok \&
van der Hulst (1998b), who find upper limits of less than 10\%.

\section{Discussion} \label{discus}
The structural parameters of the modeled disks appear to be
independent from the star formation activity.  In this section we will
discuss the appearance of the stellar disk of LSB galaxies, and the
ISM in LSB galaxies.

\subsection{Stellar disk}

LSB galaxies are flatter than normal HSB galaxies (cf Fig. 2).  From a
study of the catalog of edge-on galaxies by Karachentzev et al.\
(1993), Kudrya et al.\ (1994) concluded that galaxies become
increasingly thinner towards later Hubble types, with the very
thinnest galaxies having major over minor axis ratios of 20. Our model
galaxies have axial ratios of 15, that is, almost as flat as the
flattest galaxies in the Karachentzev et al.\ catalog.
\begin{figure}
  \epsfxsize=\hsize \hfil\epsfbox{\figuredisk/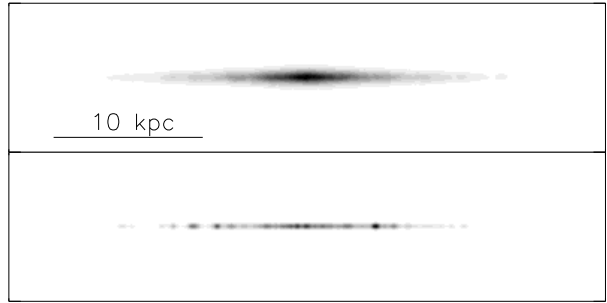}\hfil
  \caption{Edge-on views of the mass and light 
    distribution in LSB galaxy $L$. The top panel shows a linear
    grey-scale representation of the projected mass surface density.
    The bottom panel shows a grey-scale representation of the
    projected $V$-band surface brightness.  Here, the star particles
    formed during the run-time of the simulation are assigned a
    $V$-band flux according to their age following Charlot \& Bruzual
    (1993).  The contrast has been adjusted to a dynamic range of 5
    magnitudes. This view resembles the so-called ``chain galaxies''
    described by Cowie et al.\ (1995) \label{zheight}}
\end{figure}
The top panel in Fig.\ \ref{zheight} shows the edge-on view of mass
distribution in LSB galaxy $L$.

Our model LSB galaxies support claims made by Dalcanton \& Shectman
(1996) that some of the extremely flat galaxies (``chain galaxies'')
detected in medium-redshift HST images (Cowie et al.\ 1995) are simply
edge-on LSB galaxies. Cowie et al.\ present HST wide-$I$ band
observations of these chain galaxies which they describe as
``extremely narrow, linear structures ... with superposed bright
`knots'.'' A few of these galaxies were found to lie at a redshift $z
\sim 0.5$. As an $I$-band view of that redshift corresponds
approximately to a $V$-band view locally, we have attempted to
simulate a $V$-band observation of model $L$ edge-on, by assigning to
each stellar particle the $I$-band flux corresponding to its age
(Charlot \& Bruzual 1993).  This is shown in the bottom panel of
Fig.~\ref{zheight}, where the dynamic range of the picture has been
adjusted to correspond approximately with the contrast that the
observed chain galaxies have with respect to the sky-background.
Comparison with any of the galaxies in Fig.~20 of Cowie et al.\ will
show the resemblance.

How will the edge-on surface brightness distributions of LSB galaxies
compare with those of HSB galaxies of similar scale length?  Compared to
an equally large HSB galaxy, the scale height of an LSB galaxy is
smaller. Due to the low dust content in LSB galaxies (McGaugh 1994),
effects of edge-brightening will be stronger in the LSB galaxy
compared to the dust-rich HSB galaxy.  An edge-on LSB might have a
higher apparent surface brightness than an edge-on HSB galaxy of the
same size.  For example, for HSB galaxies in the Ursa Major cluster,
Tully et al.\ (1997) find a differential extinction in the $B$-band
going from face-on to edge-on of $1.8$ mag arcsec$^{-2}$. The LSB
galaxies in their sample are found to be almost transparent with a
differential extinction of only $0.1$ mag arcsec$^{-2}$.  The
almost two magnitudes extra extinction in the HSB can thus easily
compensate the few magnitudes difference in {\it intrinsic} surface
brightness, making edge-on LSB galaxies as bright in surface
brightness as edge-on HSB galaxies.

Hence it is much more difficult to distinguish between edge-on LSB and
HSB galaxies than between face-on LSB and HSB galaxies.  A smaller
number of star forming regions, extreme flatness and blue colors will
be the only easily visible distinguishing characteristics.

The reason why the stellar disks of LSB galaxies are so thin is that
they are very stable to local instabilities (Fig.\ \ref{qtoomre}). As
a consequence, bars and spiral structure, which are the most efficient
methods for heating a stellar disk (Sellwood \& Carlberg 1984), are
unlikely to develop spontaneously in LSB galaxies. Thus there is no
natural way to make the disks thick.

This explanation is supported by the rarity of barred LSB galaxies. In
the LSB galaxy catalog by Impey et al.\ (1996), only 4 percent of the
galaxies are barred, while the frequency of barred galaxies in the RC2
is some 30 percent (\cite{elm90}). Sellwood \& Wilkinson (1993) give a
fraction of 2/3 for barred HSB galaxies. The stability of LSB disks is
confirmed in a numerical study of the dynamical stability of these
systems by Mihos et al.\ (1997). Whereas we adopted the ``most likely''
solution for the mass-to-light ratio of stellar disk, these authors
considered the ``worst case'' scenario of maximum disk. However their
conclusion is the same as ours: the disks of LSB galaxies are
extremely stable.

The lack of truly LSB edge-on galaxies (Schombert et al.\ 1992) could
tell us something about the existence of very LSB galaxies. As shown
above, the presently known population of LSB galaxies, when turned
edge-on, brightens to surface brightnesses comparable to those of
similar-sized HSB galaxies. In principle, the edge-on counterparts of
the currently known LSB galaxies should thus already be in the conventional
galaxy catalogs, showing up as ``streaks on the sky''.  As truly LSB
edge-on galaxies seem to be lacking from the LSB catalogs, this
suggests that galaxies with face-on surface brightnesses $\mu_0(B) >
25$ mag arcsec$^{-2}$ are probably very rare.

\subsection{Blue \& red LSB galaxies}
Fluctuations in the SFR as shown in Fig. 3, may very well explain why
most of the LSB galaxies detected in surveys are blue.  As McGaugh
(1996) argues, the selection effects against finding red LSB galaxies
on the blue sensitive plates on which surveys have been carried out
are quite severe. It will be very hard to find LSB galaxies with
colors $B-V \simeq 1$ or redder.  So if the absence of red LSB
galaxies from the current catalogs is caused purely by these
selection effects, the blue LSB galaxies could
simply be the bursting tip of a proverbial iceberg.

In order to explain the colors of the bluest LSB galaxies in the
sample of de Blok et al.\ (1996), van den Hoek et al.\ (1997) had to
invoke bursts with a duration of between 0.5 and 5 Myr and an
amplitude between 1 and 5 $M_{\odot}$/yr.  The fluctuations found in
our simulation $L$ have an amplitude of 0.15 $M_{\odot}$/yr, with a
duration of about 20 Myr. This duration is determined by the lifetime
for OB stars and the collapse time for molecular clouds, which are
both of the order of a few $10^7$ yr. They are therefore slightly
milder than the bursts invoked by van den Hoek et al.\ (1997), but
keep in mind that the latter bursts were used to explain the {\it
  bluest} galaxies. Our models attempt to simulate an average LSB
galaxy, and it should not come as a surprise that the fluctuations we
find are milder. Given the many assumptions and uncertainties in both
modeling and observations, it is actually quite encouraging that the
parameters agree to within an order of magnitude. 

Assuming that blue LSB galaxies are currently undergoing a period of
enhanced star formation, implies that there exists a population of
{\it red, non-bursting, quiescent\/} LSB galaxies.  These should then
also be metal-poor and gas-rich, and share many of the properties of
the galaxies we have been modeling.

We can estimate the fraction of red LSB galaxies by calculating the
distribution of the birthrate parameter $b$ for simulation $L$.  The
$b$ parameter is the ratio of the present SFR over the average past
SFR and is a useful tool for studying the star formation history of
galaxies. Birthrate parameters have been determined for a large sample
of spiral galaxies by Kennicutt et al.\ (1994). The trend is that
early type galaxies have small values for $b$, thus most star
formation occurred in the past, while late type and irregular galaxies
have large values for $b$, often exceeding 1, indicating that those
galaxy are still actively forming stars, and more so than in the past.
Here we apply this analysis to simulation $L$ in order to estimate the
fractions of blue and red LSB galaxies.

\begin{figure}
  \epsfxsize=\hsize \hfil\epsfbox{\figuredisk/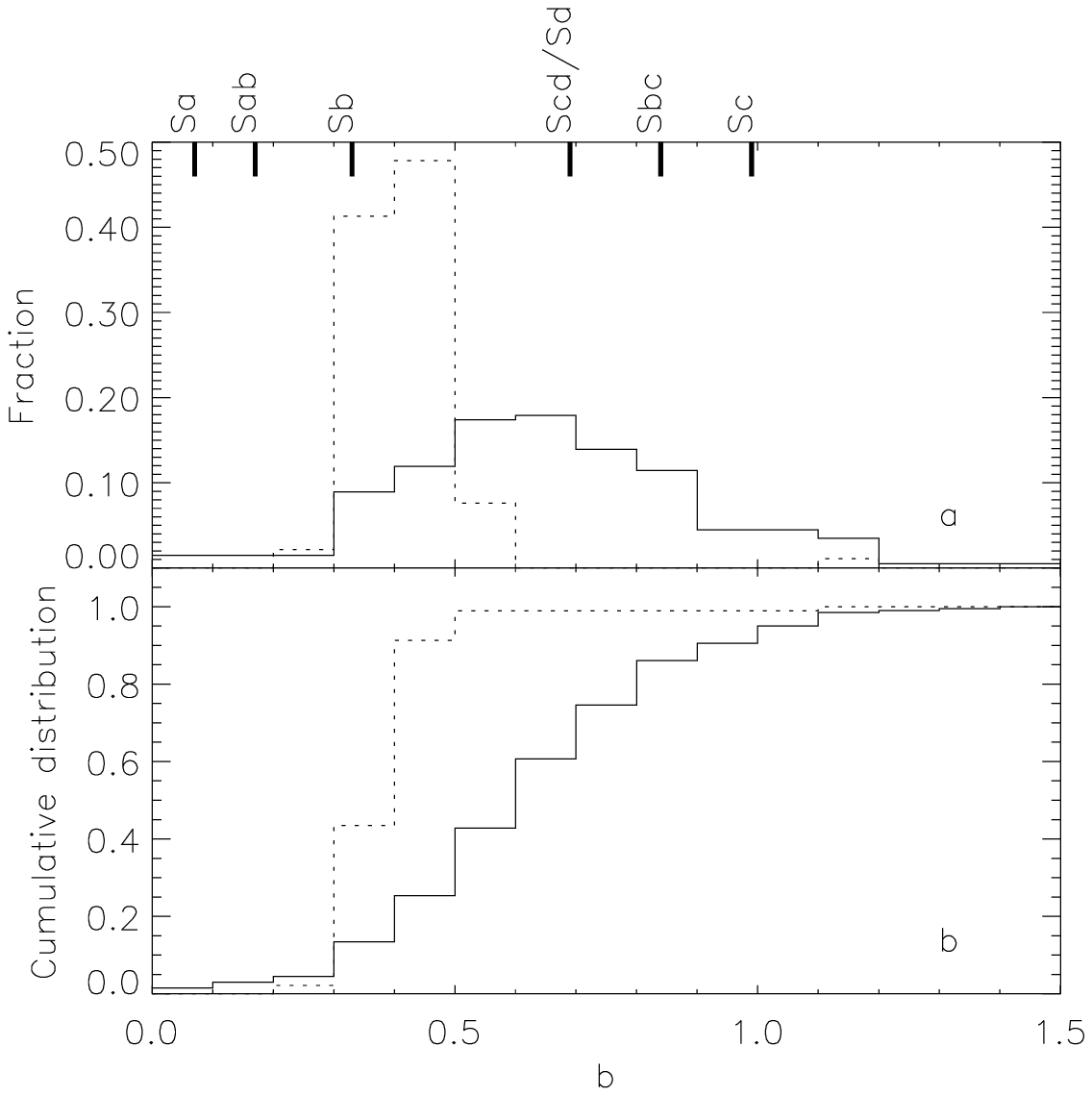}\hfil
  \caption{$(a)$ Distribution of the birthrate parameter $b$, the
  ratio of the current SFR to the average past SFR. Solid line shows
  the $b$ values derived for simulation $L$, dotted line shows the $b$
  distribution for a simulation of an Sc galaxy (\cite{ger97b}). On
  the top are average $b$ values for different types of galaxies
  (\cite{ken94}). The bottom panel shows the cumulative distribution
  for the two simulations. Less than 20\% of the $b$ values for the
  LSB galaxy are below $b=0.4$, hence we expect at most 20\% of LSB
  galaxies to be ``red''  \label{bhist}}
\end{figure}

In Fig.\ \ref{bhist}a we plot the distribution of $b$ values for
simulation $L$ (solid line), where we have followed the value of $b$
over the duration of the simulation in steps of 15 Myr. Thus if the
SFR peaks in a particular time interval, the corresponding value of
$b$ will be high. If the SFR is low in this time interval $b$ is also
low.  In total we have 200 $b$ values.

Also shown in Fig.\ \ref{bhist}a are the $b$ distribution for
a simulation of an HSB Sc galaxy (\cite{ger97b}, dotted line) and the
mean values for different galaxy types (from \cite{ken94}). Due to the
low average SFR the distribution for the LSB simulation is much
broader than the distribution for the HSB simulation, and the average
$b$ value is larger.
  
Also it is clear that the LSB galaxy has $b$ values larger than the
average for early type galaxies for most of the time. ``Classical''
LSB galaxies are blue compared to HSB galaxies, they thus have an
excess of recent star formation or equivalently a higher $b$ value.
We now define a LSB galaxy to be ``blue'' if its birthrate parameter
$b$ exceeds the average value of $b$ for a HSB late-type galaxy (see
\cite{ken94} for relations between birthrate parameter and
color). Fig.\ \ref{bhist}a shows that this requires that $b_{LSB} >
\langle b_{HSB} \rangle \approx 0.4$.

LSB galaxies that do not meet this requirement are ``red'':
non-bursting, but nevertheless still gas-rich.  From Fig.\ 
\ref{bhist}b (which shows the cumulative $b$ distribution) we can see
that over 80 percent of the fluctuations result in blue LSB galaxies.
Less than 20 percent of the fluctuations therefore results in red LSB
galaxies.

We can estimate colors of the red population using the burst models
from van den Hoek et al.\ (1997).  For example, a 5 Myr burst with an
amplitude of 3 $M_{\odot}$/yr superimposed on a 13 Gyr old
population which has undergone an exponentially decreasing SFR with a
timescale of 4 Gyr changes colors and surface brightnesses by $\Delta
(B-V) \simeq -0.4$, $\Delta (R-I) \simeq -0.2$, $\Delta B \simeq -1$
and $\Delta I \simeq -0.4$ (cf.\ Table 5 in van den Hoek et
al.). Using these values together with the measured colors of the
bluest LSB galaxies in de Blok et al. (1995), yields $B-V \sim 1$,
$R-I \sim 0.6$ and $\mu_0(B) = 24.5$ for the red population.

A recent CCD survey (O'Neill \& Bothun 1997) has picked up a class of
LSB galaxies which have $\mu_0(B) \simeq 24$ and $B-V \simeq 0.8$.  If
some of these galaxies are indeed the non-bursting counterparts of the
blue LSB galaxies, they should be metal-poor and gas-rich, and share
many of the properties of the galaxies we have been modeling.

In summary, if the blue colors found in LSB galaxies are the result of
fluctuations in the star formation rate, then this implies that the
red gas-rich LSB galaxies constitute less than 20\% of the gas-rich
LSB disk galaxies.  This does not rule out the existence of a
population of red, gas-poor LSB galaxies. These must however have had
an evolutionary history quite different from those discussed here and
possibly have consumed or expelled all their gas quite early in their
life.

\subsection{ISM}
The SFRs from both simulations are larger than the observed rate of
about $0.05\ M_{\sun}$/yr. This latter value is probably correct
within a factor of 2. Especially the high SFR of simulation $H$
($0.33\ M_{\sun}$/yr) is almost an order of magnitude larger than what
is needed and it seems impossible to reconcile this value with the
observations. Instead the global SFR is consistent with the global SFR
for equal-luminosity galaxies (e.g.\ \cite{ken83}). For instance the
Sc galaxy NGC 6503 with a maximum rotation velocity of $v=120$ km/s
has a total measured SFR of $0.4\ M_{\sun}$/yr, and a simulated SFR of
$0.35\ M_{\sun}$/yr (\cite{ger97a}, 1998).  In general the SFRs of
equal-luminosity galaxies differ by a factor 10 between LSB and HSB
galaxy (\cite{hoe97}, and Kennicutt 1983), despite the copious amounts
of gas available in the LSB galaxies (\cite{dbmh96}).

SFR$_L$ approaches the correct value and the physical reason is shown
in Fig.\ \ref{ism}. The essential information to retain from this
phase diagram is that we need a different ISM for LSB galaxies, where
the bulk of the gas is not directly available for star formation, as
it is too warm (of order $10^4$ K).  The simulations do not include
phase transitions from neutral to molecular gas, but as an estimate
for the H$_2$ mass we can consider all star forming gas ($T\la300$ K)
to be molecular. This gas represents less than 2\% of the total gas
mass.  In this respect it is interesting to note that for a small
sample of LSB galaxies CO is not detected, yielding an 
limit of M(H$_2$)/M(H\,{\sc i}) $\ll$ 30 percent (\cite{dbh98}, Schombert et
al.\ 1990). A few galaxies have upper limits less than 10\%.

As discussed in the previous section the warm ISM may be caused by a
low metallicity for the gas. Below $10^4$ K, cooling is dominated by
heavy elements like C$^+$, Si$^+$, Fe$^+$, O. If these elements are
rare then it is difficult for the gas to radiate its energy away.

Direct observational support for a low metallicity ISM in LSB galaxies
comes from oxygen abundance measurements of H\,{\sc ii} regions in LSB
galaxies.  Those studies yield metallicities of approximately 0.5
times solar metallicity (\cite{mcg94}). Measurements of the oxygen
abundance in F563-1 (de Blok \& van der Hulst 1998a) give an average
oxygen abundance of 0.15 $Z_{\odot}$ (compare with the difference in
metallicity between models $H$ and $L$).  

A point of concern is the IMF. Our premise is that the IMF is
universal, however it can differ from our adopted Salpeter IMF (see
\ref{starform}). Especially a low upper-mass cutoff could easily lead
to an underestimation of the SFR (e.g.\ an upper mass cutoff at 30
$M_{\sun}$ reduces the measured SFR by 2-3; \cite{ken83}).  However
current determinations of the IMF in external galaxies point towards a
universal IMF (\cite{ken89}; \cite{ken94}).  Furthermore, observations
suggest that massive stars {\it are} present in the H\,{\sc ii}
regions of LSB galaxies (McGaugh 1994).

\section{Conclusions} \label{conclusion}
What causes the low evolution rate for LSB galaxies? The low density
has often been invoked to explain this, since the dynamical time scales
with $1/\sqrt{\rho}$.  This scenario is exactly what is tested
in simulation $H$. The only difference with a normal HSB galaxy is the
scale length of the galaxy, which is for instance three times larger than
the scale length for the equally bright galaxy NGC 6503. The result is
striking: adopting ``standard'' values for the star formation process
results in a SFR identical to the rates of HSB galaxies.

Thus the low density in itself seems not capable of doing the job, and
we have to rely on a scarcity of heavy elements to reproduce a true
LSB galaxy. This fits in logically with the notion that stars are the
producers of these elements; the low star formation activity prevents
metal enrichment of the ISM. It implies that the SFR has been low
throughout the evolution of LSB galaxies, and that these galaxies are
``trapped'' in their current evolutionary state: low density prevents
rapid star formation, which prevents enrichment of the ISM, which
prevents cooling, resulting in a warm one-phase ISM. So although the
lack of metals is directly responsible for the low SFR, the low
density may ultimately determine the fate of LSB galaxies.

In summary we reach the following conclusions on the physical
properties of LSB galaxies.

\begin{enumerate}
\item \noindent {\bf Dominant halo:} The dominance of the dark matter halos in
  LSB galaxies (de Blok \& McGaugh 1997) results in very thin and
  stable disks.  The major over minor axis ratios of the stellar disks
  are larger than 15.  Lack of dust in LSB galaxies will result in
  edge-on LSB galaxies having apparent surface brightnesses equal to
  those of edge-on HSB galaxies of similar size.  The lack of truly
  LSB edge-on galaxies may indicate that disk galaxies with $\mu_0(B)
  > 25$ mag arcsec$^{-2}$ are rare.

\item \noindent {\bf Low metallicity:} Modeling the low SFR found in
  LSB galaxies, demands that the metallicity in the ISM must be
  approximately 0.2 solar, which is consistent with observation by
  McGaugh (1994) and de Blok \& van der Hulst (1998a).  These low
  metallicities prevent effective cooling so that the ISM in LSB
  galaxies primarily consists of warm ($\sim 10^4$ K) gas. The amount
  of cold (molecular) gas is probably less than 5\% of the H\,{\sc i}
  mass, supporting claims by Schombert et al. (1990) and de Blok \&
  van der Hulst (1998b) who derive small molecular gas fractions.
  
\item \noindent {\bf Low average SFR:} Due to the fluctuations in the
  SFR in LSB galaxies and their large contrast with the average SFR,
  the spread in colors among LSB galaxies will be larger than among
  HSB galaxies.  From the distribution of birthrate parameters in our
  simulations we deduce that, if the currently known blue {\it
  gas-rich} LSB galaxies are the most actively star forming LSB
  galaxies, they constitute over 80 percent of the total population of
  {\it gas-rich} field LSB disk galaxies.  This implies that there is
  at most an additional 20 percent of quiescent, gas-rich LSB disk
  galaxies.  This does not preclude the existence of an additional
  red, gas-poor population.  However this population must have an
  evolutionary history quite different from that described in this
  work.

\end{enumerate}

\begin{acknowledgements}
  We are much indebted to Lars Hernquist for generously providing the
  TREESPH code. We also thank Frank Briggs and Thijs van der Hulst for
  their comments on a draft of this paper. We thank the referee,
  Dr. Friedli, for his extensive comments. The research of JPEG was
  supported by the Netherlands Foundation for Research in Astronomy
  (NFRA) with financial aid from the Netherlands Organization for
  Scientific Research (NWO).
\end{acknowledgements}

\end{document}